\documentclass[prl,showpacs,twocolumn]{revtex4}
\usepackage{amssymb}
\usepackage{amsmath}
\usepackage{amsfonts}
\usepackage[]{graphicx}
\begin{document}
\bibliographystyle{prsty}
\title{Near-field microscopy with a single-photon point-like emitter: Resolution versus the aperture tip?}
\author{Aur\'{e}lien Drezet, Aur\'{e}lien Cuche, and Serge Huant}
\affiliation{Institut N\'eel UPR 2940, CNRS-University Joseph Fourier, 25 rue des Martyrs, 38000 Grenoble, France}

\date{\today}

\begin{abstract}
We discuss theoretically the concept of spatial resolution in near-field scanning optical microscopy (NSOM) in light of a recent work [Opt. Express 17 (2009) 19969] which reported on the achievement of active tips made of a single ultrasmall fluorescent nanodiamond grafted onto the apex of a substrate tip and on their validation in NSOM imaging. Since fluorescent nanodiamonds tend to decrease steadily in size, we assimilate a nanodiamond-based tip to a point-like single photon source and compare its ultimate resolution with that offered by standard metal-coated aperture NSOM tips. We demonstrate both classically and quantum mechanically that NSOM based on a point-like tip has a resolving power that is only limited by the scan height over the imaged system whereas the aperture-tip resolution depends critically on both the scan height and aperture diameter. This is a consequence of the complex distribution of the electromagnetic field around the aperture that tends to artificially duplicate the imaged objects. We show that the point-like tip does not suffer from this ``squint'' and that it rapidly approaches its ultimate resolution in the near-field as soon its scan height falls below the distance between the two nano-objects to be resolved.
\end{abstract}

\maketitle


\section{Introduction}
Since its birth in the early 80's~\cite{Pohl}, the near-field scanning optical microscope, or NSOM~\cite{Courjon:2003},  became a versatile tool for imaging optical properties of systems requiring a very high spatial resolution in the nanometer range~\cite{Novotny:2006,Greffet}. Yet one of the most fundamental issue with NSOM concerns the optical resolution available  with a given system. Current systems based on aperture-NSOM~\cite{Pohl} with a hole at the apex of a metal-coated conical tip are fundamentally limited by the size of the optical hole~\cite{Betzig,Hulst,Drezet2003a}. In order to improve the optical resolution of NSOM and therefore overcome this limitation one could ideally use a point-like emitting source. Recently, inspired by the pioneer work by Michaelis \emph{et al.}~\cite{Mich} who used a fluorescent single molecule at low temperature as basis for a NSOM, we developed a new high-resolution NSOM tip using a nitrogen-vacancy (NV) color-center in a diamond nanocrystal as a scanning point-like light source~\cite{Cuche}. In this active tip the 20 nm nanocrystal is glued \emph{in situ} at the apex of an etched optical fiber probe. Importantly, the NV center acts as a photostable (no blinking, no bleaching) single-photon source working at room temperature~\cite{Poizat,Yannick}. As such, the NV-center based tip proves superior to quantum-dot based tips~\cite{Chicolas}, which suffer from unsufficient photostability~\cite{YS2006}, and to insulating-nanoparticle based tips,  which, despite remarkable photostability, cannot (yet) reach the single-photon emission rate~\cite{Nanotec2009}. Therefore, the nanodiamond-based NSOM probe opens large avenues for microscopy and quantum optics in the near-field regime.\\
In the present work we study theoretically the potential resolution offered by the nanodiamond-based NSOM tip that we assimilate to a point-like emitter and we compare it with the resolution of a standard aperture-tip. For this we will first describe simple electromagnetic models for both the point-like (sections II) and the aperture (section III) optical probes. In section II we return to and justify some of our assumptions concerning the photon emission by the NV-center based NSOM and we discuss in more detail the impact of quantum optics on the optical resolution of this system. Finally, in section IV we apply and compare these models for simulating an image acquisition during a scan over fluorescent test systems.

\section{Near-field optical model of the single-photon tip}
\subsection{The point like source in classical optics}
The concept of optical resolution is at the heart of any photonic microscopy~\cite{Barchiesi}. As it is well known the resolving power of classical microscopy, that is the ability to separate spatially two point-like objects, is limited to spatial separations $d$ larger than half the illumination wavelength: $d\gtrsim\lambda/2$. Fundamentally this is due to the fact that far-field microscopy collects only propagative waves for imaging. In other words, the microscope acts as a low-pass filter for the incident field plane wave expansion since only wavevectors with planar components satisfying $k_x^2+k_y^2\leq (2\pi/\lambda)^2$ contribute to the image reconstruction. This leads to a  Fourier-Heisenberg like uncertainty relation $\Delta k_x\cdot \Delta x \gtrsim 1/2$ where $\Delta k_x\sim 1/\lambda$  and $\Delta x\sim d$ and thus to the Rayleigh-Abbe criterion~\cite{Born,Novotny:2006}: $d\gtrsim\lambda/2$.\\
Near- field optical microscopes circumvent this limitation by including the neglected components with  $k_x^2+k_y^2\geq (2\pi/\lambda)^2$ in the imaging procedure~\cite{Courjon:2003,Novotny:2006,Greffet}. More precisely, in NSOM in the illumination configuration~\cite{Pohl} a tip-probe is brought into the vicinity of the structure to be observed, i.e., at a distance much smaller than the optical illumination wavelength $z\sim \lambda /30 \simeq  20$ nm. Evanescent components of the emitted light (the so-called near-field contributions) are used to resolve spatially the objects on the surface. A different way to see this is to use a multipolar expansion of the field around the tip apex. It is widely accepted that this field is well described by the dominant (electric and magnetic) dipolar terms. For example, the electromagnetic field $\mathbf{E}$,  $\mathbf{B}$ of such an (harmonic) electric point-like dipole term $\mathbf{P}_\omega e^{-i\omega t}$ calculated at position $\mathbf{r}$ and time $t$ is given by
\begin{eqnarray}
\mathbf{E}(\mathbf{r},t)=\mathbf{G}^0_\omega(\mathbf{r},\mathbf{r}_0)\cdot
\mathbf{P}_\omega e^{-i\omega t};&
ik\mathbf{B}=\boldsymbol{\nabla}\times\mathbf{E}
\end{eqnarray}
where $\mathbf{r}_0$ is the location of the point dipole,
$k=\omega/c$, and $\mathbf{G}^0_\omega(\mathbf{r},\mathbf{r}_0)$ is
the dyadic Green propagator of the dipole \cite{Novotny:2006,Girard} in vacuum which reads $\mathbf{G}^0_\omega(\mathbf{r},\mathbf{r}_0)=[(\omega/c)^2\mathbf{I}+\boldsymbol{\nabla}\boldsymbol{\nabla}]G^{0}_{\omega}(|\mathbf{r}-\mathbf{r}_0|)=\boldsymbol{\nabla}\times\boldsymbol{\nabla}\times (\mathbf{I}G^{0}_{\omega}(|\mathbf{r}-\mathbf{r}_0|))$ with $G^{0}_{\omega}(|\mathbf{r}-\mathbf{r}_0|)=e^{ik|\mathbf{r}-\mathbf{r}_0|}/(4\pi|\mathbf{r}-\mathbf{r}_0|)$ and $\mathbf{I}$ the diagonal  Kronecker tensor. The dipolar field can explicitly be written~\cite{Jackson}  (in
Heaviside-Lorentz units) as
\begin{equation}\left\{
\begin{array}{ll}
\mathbf{B}=k^{2}\left(\frac{\mathbf{R}}{R}\times \mathbf{P}_\omega\right)\frac{e^{ik(R-ct)}}{4\pi R}\left(1-\frac{1}{ikR}\right)& \\
&\\
\mathbf{E}=k^{2}\left(\frac{\mathbf{R}}{R}\times \mathbf{P}_\omega\right)\times \frac{\mathbf{R}}{R}\frac{e^{ik(R-ct)}}{4\pi R} & \\ \\+[3\frac{\mathbf{R}}{R}\left(\frac{\mathbf{R}}{R}\cdot\mathbf{P}_\omega \right)-\mathbf{P}_\omega ]\left(\frac{1}{4\pi R^{3}}-\frac{ik}{4\pi R^{2}}\right)e^{ik(R-ct)}&
\end{array}\right.
\end{equation}
where  $\mathbf{R}=\mathbf{r}- \mathbf{r}_0$.\\
 It is easily seen that in the near-field regime  where $kR\ll 1$ we have the asymptotic behavior:
\begin{equation}
\left\{\begin{array}{ll}
\mathbf{B}=ik\left(\frac{\mathbf{R}}{R}\times \mathbf{P}_\omega\right)\frac{1}{4\pi R^{2}}&\\&\\
\mathbf{E}= [3\frac{\mathbf{R}}{R}\left(\frac{\mathbf{R}}{R}\cdot\mathbf{P}_\omega \right)-\mathbf{P}_\omega ]\frac{1}{4\pi R^{3}} \end{array}\right.
\end{equation}
This electric near-field, which is clearly highly localized around the dipole (with $|\mathbf{E}|\propto 1/R^3$), is actually
responsible for the super resolution offered in the near-field by the ideal point-like source considered here.\\
In the nanodiamond-based NSOM probe~\cite{Cuche}, the optical behavior originates from the point-like transition dipole associated with the radiative transition of the NV color-center. We here neglect the finite extension of the host nanocrystal since current trends in nanodiamond processing indicate that it can be diminished down to a few nms only~\cite{boudou,small2009}. Therefore, we expect the point-like model outlined above to apply to this tip and we will indeed describe the nanodiamond-based tip as a single point-like dipole.

\subsection{Optical resolution and quantum optics}
As discussed in the introduction, the NSOM experiments reported in \cite{Cuche} involves a single nanodiamond containing one and only one fluorescent quantum emitter i.e. a NV center. This point is of general importance for the present discussion since such a quantum object acts like a single-photon source emitting photon one by one in the typical lifetime of the (two-level) optical transition $\Gamma^{-1}\simeq 10$ ns~\cite{Cuche}. In particular, we already pointed out \cite{Cuche} that such a NV-based NSOM constitutes ultimately a near-field scanning single-photon microscope. To justify this point it is sufficient to note that $\Gamma^{-1}$ corresponds to a length scale of 3 m, i.e., to a size which is by
order of magnitudes larger than any device scales in nano and microphotonics. Therefore, we can fairly
state that during this lifetime unit scale there is only one photon created by
the quantum emitter in the whole imaged photonic system.\\
The question which naturally arises concerns the implication of
quantum optics on the concept of optical resolution since we know
that in quantum optics electromagnetic field components are
associated with non commuting operators. Actually,
Heisenberg~\cite{Heisenberg} already attempted in the 1930's to
introduce some limitations in the measurability of electromagnetic
field quantities in a similar way to that he used for
deriving the well known uncertainty relation for complementary
kinematic variables such as position and momentum. The
argumentation was however criticized by Landau and
Peierls~\cite{Landau} who demonstrated that further restrictions
in quantum field theory were necessary. This point was later
clarified by Bohr and Rosenfeld in a well known paper~\cite{Bohr}
where they showed that the method chosen for measuring the fields
(and in particular the space and time averaging process) is the
key issue~\cite{Bohr,Bohr2,Darrigol}. Recently, the problem has
gained a renewed interest in the context of
nano-optics~\cite{Barnes,Dereux} (see also
\cite{Birula,Birula2,Keller} for important related topics). More
precisely, starting from the Heisenberg uncertainty
relation~\cite{Heisenberg} which links through a circular
permutation the components of the electric field  $\overline{E}_i$
and magnetic field $\overline{B}_j$ averaged in a cubic volume
$\delta L^3$
\begin{equation}
\delta \overline{E}_i\cdot\delta \overline{B}_j =\delta \overline{E}_j\cdot\delta \overline{B}_i\gtrsim \frac{\hbar c}{2\delta L^4},
\end{equation}
it was deduced that in the context of NSOM, measurements of the
fields and of the energy should become
uncertain~\cite{Barnes,Dereux} (see however~\cite{foot}). This is
because volumes much smaller than the optical wavelength (cubed)
are probed. As any NSOM experiment ultimately involves a detection
of photons one could therefore suspect a supplementary and purely
quantum alteration of the optical resolution.  Here, we show
however that Eq.~4 does not actually limit the resolution of the
NV-based NSOM. To see why this is the case we must go back to the
dynamics of the NV center assimilated with a two-level fluorescent
system. We have in particular to justify some assumptions done in
the previous subsection about the use of classical formulas for the
electromagnetic field generated by the NV center.
\subsection{Spontaneous emission}
As a first approach it is enough to consider a model of the photon emission based on the Wigner-Weisskopf theory for a two-level point-like radiating system~\cite{Scully,Sipe}. Rigorously speaking we would need to use quantum electrodynamics for a transition dipole located in a inhomogeneous dispersive and dissipative dielectric (or metallic) environment. This would imply including the degree of freedom of the polarizable medium in the hamiltonian treatment of the problem. For the present purpose we choose a simplified configuration and consider the emission in vacuum only. Following this model the two-level system initially in the excited quantum state $|e\rangle$ (with energy $E_e$) at $t=0$ spontaneously relaxes with a typical rate $\Gamma$ to the ground state $|g\rangle$ (with energy $E_g$). At any time $t\geq0$ the whole quantum system involving photon and emitter is in an entangled state which can formally be written
\begin{equation}
|\Psi(t)\rangle=c_e(t)|e\rangle\otimes|0\rangle+ |g\rangle\otimes|\gamma(t)\rangle,
\end{equation} where  $|0\rangle$, $|\gamma(t)\rangle$ are respectively the vacuum electromagnetic state and the single-photon state at time $t$.  Additionally, $c_{e}(t)$ is a complex valued amplitude coefficient depending on time and we have $c_{e}(t)=\Theta(t)e^{-i\omega_{eg}t-\Gamma t/2+i\Delta t}$  where $\hbar\omega_{eg}=E_e-E_g$, $\Delta$ is the Lamb-shift, and $\Theta(t)$ is the Heaviside unit step function. The dissipation constant is given by the usual formula $\hbar\Gamma=(\omega_{eg}/c)^3|\boldsymbol{\mu}_{ge}|^2/(3\pi)$ where $\boldsymbol{\mu}_{ge}=\langle g|\boldsymbol{\mathbb{\mu}}|e\rangle$ is the transition dipole amplitude. This naturally leads to the usual exponential probability rule for ``non-decay'' $p_e(t)=|c_e|^2= e^{-\Gamma t}$. We also define a ground state $|G(t)\rangle=|g\rangle\otimes|0\rangle$ such as $|G(t)\rangle=|G(0)\rangle$ as expected (i.~e., the ground state is stable).
Now, using the total Power-Zienau Hamiltonian operator for the
interacting system field plus a two-level system in the dipolar
approximation~\cite{Sipe,PowerT,Tannoudji}
\begin{eqnarray}
\hat{H}(t)\simeq \int d^{3}\mathbf{r}\frac{\mathbf{\hat{D}}(\mathbf{r},t)^2+\mathbf{\hat{B}}(\mathbf{r},t)^2}{2}\nonumber\\+E_g|g\rangle\langle g|_{(t)}+E_e|e\rangle\langle e|_{(t)}
-\boldsymbol{\hat{\mu}}(t)\cdot\mathbf{\hat{D}}(\mathbf{r}_0,t).
\end{eqnarray}
We obtain (in the Heisenberg picture) the quantum-Maxwell equations \cite{Sipe}:
\begin{eqnarray}
0=\boldsymbol{\boldsymbol{\nabla}}\cdot\mathbf{\hat{D}}(\mathbf{r},t),\nonumber\\0=\boldsymbol{\boldsymbol{\nabla}}\cdot\mathbf{\hat{B}}(\mathbf{r},t),\nonumber\\
\boldsymbol{\boldsymbol{\nabla}}\times\mathbf{\hat{B}}(\mathbf{r},t)=\frac{1}{c}\frac{\partial \mathbf{\hat{D}}(\mathbf{r},t)}{\partial t},\nonumber\\
\boldsymbol{\boldsymbol{\nabla}}\times\mathbf{\hat{D}}(\mathbf{r},t)=-\frac{1}{c}\frac{\partial \mathbf{\hat{B}}(\mathbf{r},t)}{\partial t} +\boldsymbol{\boldsymbol{\nabla}}\times\mathbf{\hat{P}}(\mathbf{r},t)
\end{eqnarray}
where $\mathbf{\hat{P}}(\mathbf{r},t)=\boldsymbol{\hat{\mu}}(t)\delta^{3}(\mathbf{r}-\mathbf{r}_0)$ is the dipole moment Heisenberg operator for the point-like source located at $\mathbf{r}_0$. $\mathbf{\hat{D}}(\mathbf{r},t)=\mathbf{\hat{E}}(\mathbf{r},t)+\mathbf{\hat{P}}(\mathbf{r},t)$, and $\mathbf{\hat{E}}(\mathbf{r},t),\mathbf{\hat{B}}(\mathbf{r},t)$ are electromagnetic field Heisenberg operators.\\
In the context of Glauber theory of photo-detection~\cite{Glauber} the most relevant quantities are the transition amplitudes
\begin{equation}
\mathbf{\mathbb{D}}(\mathbf{r},t)=\langle G(0)|\mathbf{\hat{D}}(\mathbf{r},t)|\Psi(0)\rangle=\langle 0|\mathbf{\hat{D}}(\mathbf{r})^{(+)}|\gamma(t)\rangle,...
\end{equation} with similar mathematical expressions for  $\mathbf{\mathbb{E}}$ and $\mathbf{\mathbb{B}}$ ($\mathbf{\hat{X}}(\mathbf{r})^{(+)}$ is a destruction operator in the Schr\"{o}dinger representation, i.e., the positive frequency part operator~\cite{Glauber,Scully}). We also introduce the transition dipole volume density
\begin{eqnarray}
\mathbf{\mathbb{P}}(\mathbf{r},t)=\langle G(0)|\mathbf{\hat{P}}(\mathbf{r},t)|\Psi(0)\rangle=c_e(t)\boldsymbol{\mu}_{ge}\delta^{3}(\mathbf{r}-\mathbf{r}_0).
\end{eqnarray}
All these (complex valued) quantities obey rigorously to the same Maxwell equations as defined in Eq.~7 and therefore  the space time evolution of $\mathbf{\mathbb{E}}$ and $\mathbf{\mathbb{B}}$ is identical to the one associated with the classical (complex) Maxwell field  generated by a classical oscillating and damped point-like dipole $\boldsymbol{\mathbb{\mu}}(t)=c_e(t)\boldsymbol{\mu}_{ge}$~\cite{DrezetFound,DrezetPRA}. In particular the electric field is given by
\begin{equation}
\mathbf{\mathbb{E}}(\mathbf{r},t)=\int d\omega \mathbf{G}^0_\omega(\mathbf{r},\mathbf{r}_0)\cdot
\boldsymbol{\mu}_\omega e^{-i\omega t}
\end{equation}
where
\begin{eqnarray}
\boldsymbol{\mu}_\omega=\frac{\boldsymbol{\mu}_{ge}}{2\pi}\frac{i}{\omega-\omega_0}
\end{eqnarray}
is the Fourier transform of $\boldsymbol{\mathbb{\mu}}(t)$ and $\omega_0=\omega_{eg}-\Delta-i\Gamma/2$. Using the definition $\mathbf{G}^0_\omega(\mathbf{r},\mathbf{r}_0)=\boldsymbol{\nabla}\times\boldsymbol{\nabla}\times (\mathbf{I}G^{0}_{\omega}(|\mathbf{r}-\mathbf{r}_0|))$ we then get
\begin{eqnarray}
\mathbf{\mathbb{E}}(\mathbf{r},t)=\boldsymbol{\nabla}\times\boldsymbol{\nabla}\times \left\{\frac{\boldsymbol{\mu}_{ge}}{4\pi R} \right. \nonumber\\
\left.\int \frac{d\omega}{2\pi} e^{-i\omega(t-R/c)}\frac{i}{\omega-\omega_0}\right\}\nonumber\\
=\boldsymbol{\nabla}\times\boldsymbol{\nabla}\times \left(\frac{\boldsymbol{\mu}_{ge}}{4\pi R}\Theta(t-R/c)e^{-i\omega_{0}(t-R/c)}\right).\nonumber\\
\end{eqnarray} This retarded field reads explicitly
\begin{eqnarray}
\mathbf{\mathbb{E}}(\mathbf{r},t)=\left\{(\omega_{0}/c)^2\left(\frac{\mathbf{R}}{R}\times \frac{\boldsymbol{\mu}_{ge}}{4\pi R}\right)\times \frac{\mathbf{R}}{R}\right. \nonumber \\+\left. [3\frac{\mathbf{R}}{R}\left(\frac{\mathbf{R}}{R}\cdot\boldsymbol{\mu}_{ge}\right)-\boldsymbol{\mu}_{ge} ]\left(\frac{1}{4\pi R^{3}}-\frac{i\omega_{0}/c}{4\pi R^{2}}\right)\right\}\nonumber\\ \cdot e^{-i\omega_{0}(t-R/c)}\Theta(t-R/c)\nonumber\\
\end{eqnarray}  which is mathematically equivalent to the classical formula Eq.~2 after substituting $k\rightarrow \omega_{0}/c$ and multiplying by the step function.
We can therefore state that in our model there is
no special feature of the resolution involving
quantum optics which cannot already be explained using classical
or semiclassical physical arguments.
Remark that our assumptions concerning the dipole in vacuum are not fundamental for our
reasoning. Indeed, in presence of a dielectric medium the
classical analogy is still valid and one only  must substitute the
dyadic Green function
$\mathbf{G}^0_\omega(\mathbf{r},\mathbf{r}_0)$ for a dipole in
vacuum by the total Green function
$\mathbf{G}^{\textrm{total}}_\omega(\mathbf{r},\mathbf{r}_0)$
including the reflectivity of the surrounding environment
~\cite{Wylie,Girard}. This will be used in section IV.

\subsection{Stationary regime}
It is important to notice that we have considered in the previous
section the non-stationary regime of spontaneous emission only
because the involved fields ($\mathbf{\mathbb{E}}$,
$\mathbf{\mathbb{B}}$...) have a direct analogy with the classical
fields of Maxwell equations. In other words, they act like first
quantized wave functions for the single-photon state
\cite{Sipe,Scully,DrezetFound} and lead, therefore, to a simple
``classical-like'' physical interpretation of results obtained,
however, in the framework of quantum electrodynamics.
Nevertheless, the limitation to the transitory regime is not
fundamental and if we consider the stationary regime in which the
pumping rate of the (continuous and coherent) excitation laser
source (like in e.g. ref.~\cite{Cuche}) is taken into account
explicitly, the deductions will actually be very similar. To see
that, we first integrate formally the quantum Maxwell equations
Eq.~7 for the point-like dipole source and we obtain (see for
example \cite{Kimble,Tannoudji,Milonni})
\begin{eqnarray}
\mathbf{\hat{E}}_{\textrm{source}}(\mathbf{r},t)=-\left(\frac{\mathbf{R}}{R}\times \ddot{\hat{\boldsymbol{\mu}}}(t-R/c)\right)\times \frac{\mathbf{R}}{R}\frac{1}{c^2 4\pi R} \nonumber\\ +[3\frac{\mathbf{R}}{R}\left(\frac{\mathbf{R}}{R}\cdot\hat{\boldsymbol{\mu}}(t-R/c)\right)-\hat{\boldsymbol{\mu}}(t-R/c) ]\frac{1}{4\pi R^{3}} \nonumber\\
+[3\frac{\mathbf{R}}{R}\left(\frac{\mathbf{R}}{R}\cdot\dot{\hat{\boldsymbol{\mu}}}(t-R/c) \right)-\dot{\hat{\boldsymbol{\mu}}}(t-R/c) ]\frac{1}{c4\pi R^{2}}
\end{eqnarray} which is identical to the classical formula.  As before we are interested
in the detection of emitted photons by a fluorescent nanosphere
located close to the dipole $\boldsymbol{\mu}$  and therefore  to
the calculation of photon observables like the intensity
$I(\mathbf{r},t):=
\langle\mathbf{\hat{E}}^{(-)}_{\textrm{source}}(\mathbf{r},t)\mathbf{\hat{E}}^{(+)}_{\textrm{source}}(\mathbf{r},t)\rangle$
\cite{Glauber}. Ultimately such calculations reduce to the knowledge
of dipole observables like
$\langle\boldsymbol{\hat{\mu}}^{(-)}_i\boldsymbol{\hat{\mu}}^{(+)}_j\rangle$
$([i,j]=1,2,3)$ evaluated at a retarded time $t-R/c$. The dynamics
of the NV-center observables is governed by the evolution of the
reduced density operator $\hat{\sigma}(t)$ , which is a solution of
the well-established optical Bloch equations for a two-level system
coupled to a coherent monochromatic light excitation. Such a
coherent excitation  state $|\alpha\rangle$ at frequency $\omega_L$
is here characterized by the electric field $\langle
\alpha|\mathbf{\hat{E}}(\mathbf{r}_0,t)|\alpha \rangle=\mathbf{E}_0
\cos{(\omega_L t+\phi_0)}$ and the (complex valued) Rabi Frequency
$\Omega_R=\boldsymbol{\mu}_{ge}\cdot\mathbf{E}_0/\hbar$.  After an
unitary transformation the coherent state can be transformed into
the vacuum state $|0\rangle$  and the laser field becomes a
C-number~\cite{Tannoudji}. In other words it is equivalent to
consider the semiclassical optical Bloch equations where the
excitation is classical and the emission is
quantum~\cite{Tannoudji}. In the rotating wave and
adiabatic~\cite{remarque} approximations  we get the rate
equations~\cite{Tannoudji,Cuche}
\begin{eqnarray}
\frac{d}{d\tau}\left(\begin{array}{c}
\sigma_{ee}\\ \sigma_{gg}
\end{array}\right)\simeq\left(\begin{array}{cc}
-\Gamma'-\Gamma & +\Gamma'\\ \Gamma'+\Gamma & -\Gamma'
\end{array}\right)\cdot\left(\begin{array}{c}
\sigma_{ee}\\ \sigma_{gg}
\end{array}\right),
\end{eqnarray} where $\Gamma$ is as before the spontaneous emission constant and
\begin{equation}
\Gamma'=\frac{|\Omega_R|^2}{2}\frac{\Gamma/2}{(\Gamma/2)^2+(\omega_L-\omega_{eg}+\Delta)^2}
\end{equation} is the pumping rate (we here neglected phonon like dephasing $\gamma$ an approximation valids at low temperature). In the stationary regime the coherence term $\sigma_{eg}=\sigma_{ge}^{\ast}$
satisfies the relation $|\sigma_{eg}|^2\simeq\sigma_{ee}$ with
$\sigma_{ee}=1-\sigma_{gg}=\Gamma'/(2\Gamma'+\Gamma)$ (an
approximation valid under the  weak-field condition
$\Gamma'<<\Gamma/2$). Under the rotating wave
approximation~\cite{Kimble,Tannoudji} this allows us to derive the
photo-detection signal
\begin{eqnarray}
I_e(\mathbf{r},t)=|\mathbf{G}^0_{\omega_L}(\mathbf{r},\mathbf{r}_0)\cdot
\boldsymbol{\mu}_{ge}|^2\cdot\sigma_{ee}\end{eqnarray}
that is
\begin{eqnarray}
I_e(\mathbf{r},t)=|\mathbf{G}^0_{\omega_L}(\mathbf{r},\mathbf{r}_0)\cdot\boldsymbol{\mu}_{ge}|^2 |\sigma_{eg}|^2:=|\mathbf{E}|^2.\end{eqnarray}
Like for the transitory regime of spontaneous emission (Eq.~13) this formula is fundamentally classical in its form.
Eq.~18 is indeed the intensity of the electric field $\mathbf{E}$ defined in Eq.~2 for a dipole $\boldsymbol{\mu}_{ge}\sigma_{eg}$
driven at the laser frequency $\omega_L$. Consequently the deductions concerning the spatial resolution obtained with a classical and oscillating
point-like dipole are kept unchanged within a genuine photon-matter quantum dynamics approach (this classical analogy will be developed  in section IV).
We point out that while the imaged fluorescent molecules are electric dipolar in our analysis, this does not constitute a fundamental limitation whatsoever.
Actually if the scanned objects are sensitive to the magnetic field produced by the NV-based tip the signal would be proportional to
\begin{eqnarray}
I_m(\mathbf{r},t)=(c/\omega_L)^2|\boldsymbol{\nabla}\times[\mathbf{G}^0_{\omega_L}(\mathbf{r},\mathbf{r}_0)\cdot\boldsymbol{\mu}_{ge}]|^2 |\sigma_{eg}|^2
\end{eqnarray} i.e. $I_m(\mathbf{r},t):=|\mathbf{B}|^2$ in agreement with the classical formulas Eqs.~1, 2. More generally nothing prevents us to consider
detectors sensitive to $E_i$ and $B_j$ ($[i,j]=1,2,3$). Following Glauber theory of photo-detection this would lead to the calculation of signals having the
form  $\langle (a^\ast\hat{E}^{(-)}_i(\mathbf{r},t)+b^\ast \hat{B}^{(-)}_j(\mathbf{r},t))(a\hat{E}^{(+)}_i(\mathbf{r},t)+b \hat{B}^{(+)}_j(\mathbf{r},t))\rangle$
which again results in a classical-like formula $|aE_i+bB_j|^2$. Therefore we can conclude that there is no fundamental limitation here concerning the measurability of
electric and magnetic fields if those fields are defined by Eqs.~1, 2 and 18, 19.

\section{The aperture NSOM tip}

As it is well known, actual near-field microscopes are generally far from the point like configuration described in the previous section. Indeed, a typical NSOM uses a metal (aluminum) coating along the conical part of the tip for eliminating the background light leaking from the dielectric glass tip onto the sample. Without this opaque coating the resolution of NSOM is diffraction limited to dimensions at best~\cite{Bozhe} equal to $d\sim \lambda-\lambda/2$. At the apex of the tip a small aperture in the metal coating  (with typical diameter $2a\simeq 50-100$ nm) is created  which allows confining the optical excitation in a very small volume of typical size $a^3$.   The resolution of such tips is highly enhanced~\cite{Courjon:2003,Novotny:2006} and goes up to dimensions which are essentially diameter limited: $d\simeq 2a$. We will come back to this point paper later in the paper.\\
While the far-field generated by the aperture (i.e., for $kR\gg1$) is still in a good approximation described by a dipolar behavior, as confirmed experimentally~\cite{Karrai} and theoretically~\cite{Drezet2000b, Drezet2000c},  the near-field deviates strongly from this simplified assumption. This has been  confirmed in several studies~\cite{Betzig,Drezet2003a,Betzigb,Hulk,Drezet2003b,Brun2003}). Therefore, another approach must be developed.\\
Historically the first model used for describing an aperture near-field probe is the one given  by Rayleigh, Bethe, and Bouwkamp
for the transmission by a subwavelength circular aperture in a infinitely thin and perfectly conducting flat screen~\cite{Bethe,Bouwkamp}. However, because of important differences between the geometry considered for the aperture-NSOM tip on the one hand and the ideal flat and thin metal screen of the ``Bethe'' model on the other hand~\cite{Drezet2000b, Drezet2000c} we developed some time ago a different approach~\cite{Drezet2003a,Drezet2003b}, which takes into account the conical geometry of the optical tip  and also the various field symmetries. In this model the sources of charge and current around the aperture are considered explicitly in order to calculate the field they generate. Additionally, we compared the electric field generated by such a model current and charge distribution with the one predicted using a simpler approach, called ``ring-like model'' hereafter, where the sources of electric current and charge are confined along the annular rim of the aperture~\cite{Drezet2003b}. The very good agreement between both models, which was also confirmed independently by Antosiewicz and Szoplik by comparing the ring-like model predictions with finite difference time domain calculations~\cite{anto}, allows us to use only the ring-like model for the present work.\\
The ring-like model postulates that the electric charge volume density along the ring of radius $a$ in the $z=0$ plane is defined as
\begin{eqnarray}
\eta_\omega(\mathbf{r})=\eta_\omega(\rho,\phi,z)=\sigma_0\cos{(\phi)}\delta(\rho-a)\delta(z)
\end{eqnarray}
where $[\rho,\phi,z]$ are cylindrical coordinates and $\sigma_0$
is a charge density per length unit of the ring. The $\cos{(\phi)}$
is reminiscent of the $x$-electric polarization of the incident
field propagating in the NSOM fiber tip~\cite{Drezet2003a}. The
far-field generated by such a ring-like source is to a good
approximation electric-dipolar. Starting from the definition of the
electric dipole:
\begin{equation}\mathbf{P}_\omega:=i\int\frac{\mathbf{J}_\omega(\mathbf{r})}{\omega}
d^3\mathbf{r}=\int \mathbf{r}\eta_\omega(\mathbf{r})
d^3\mathbf{r}\label{bigdipole}\end{equation} with
$\mathbf{J}_\omega$ the electric current volume  density along
the rim, using then the charge conservation
$\boldsymbol{\nabla}\cdot\mathbf{J}_\omega=i\omega\eta_\omega$ and considering
a current propagating along the $\phi$ direction, we deduce that
\begin{equation}J_{\phi,\omega}(\mathbf{r})=i\omega
a\sigma_0\sin{(\phi)}\delta(\rho-a)\delta(z).\end{equation}
Therefore the electric dipole is
\begin{equation}\mathbf{P}_\omega=\sigma_0a^2\pi\mathbf{e}_x\end{equation}
which is aligned along the incident polarization direction. We
also point out that in this model the magnetic dipole reads
$\mathbf{M}_\omega=(1/2)\cdot\int
\mathbf{r}\times\mathbf{J}_\omega(\mathbf{r}) d^3\mathbf{r}/c=0$ , which is in agreement
with the fact that the oscillations of the electric current do not constitute a complete circulation of charges around the ring due to
the presence of the sine term in Eq.~22.  The electromagnetic
field generated by such a distribution of current and charge is
obtained by elementary integration of Eq.~1:
\begin{eqnarray}
\mathbf{E}(\mathbf{r},t)=\int
\mathbf{G}^0_\omega(\mathbf{r},\mathbf{r}')\cdot
\frac{i\mathbf{J}_\omega(\mathbf{r}')}{\omega} e^{-i\omega t}d^3\mathbf{r}';&
ik\mathbf{B}=\boldsymbol{\nabla}\times\mathbf{E}\nonumber\\
\end{eqnarray} and it indeed approaches the field generated by the dipole $\mathbf{P}_\omega$ given at large distance by Eq.~\ref{bigdipole}.\\
A few additional remarks about this model are here useful.
First, the fact that the current has only a $\phi$ component and
not a $\rho$ or a $z$ component can be justified on symmetry
ground by referring to mathematical results by Sommerfeld,
Bouwkamp and others~\cite{Bouwkamp} showing that in a diffraction
problem the dominant contribution of $\mathbf{J}_\omega$ near an
edge or a corner must necessarily be the components parallel to the
rim. Second, it should also be pointed out that the use of a ring-like
model is not new in diffraction theory since it was introduced
at the beginning of the XX$^{th}$ century by Love, Kottler,
Stratton and others \cite{Bouwkamp}. However, there was to our
knowledge no use of such a model for describing the NSOM tip
emission prior to our own work.\\ Additionally, it should be observed that
in complete analogy with the old work by Stratton and Chu
(reviewed by Bouwkamp~\cite{Bouwkamp}) we can also include in
the present description a fictitious density of ``magnetic charge''
$\gamma$ and magnetic current $\mathbf{K}$ along the rim in order
to introduce a net magnetic dipole in the formalism. The formulas
for the dipole and fields generated by such  magnetic distributions
are obtained by analogy with the electric ones (i.e.,
Eqs.~1-5 and 20-24) and using the direct substitutions
$\mathbf{E}\rightarrow\mathbf{B}$,
$\mathbf{B}\rightarrow-\mathbf{E}$, $\eta\rightarrow\gamma$,
$\mathbf{J}\rightarrow\mathbf{K}$ which also imply
$\mathbf{P}\rightarrow\mathbf{M}$. The possibility to introduce
magnetic distributions and dipoles is justified in order to
explain the empirical observations by Oberm\"{u}ller \emph{et al.}~\cite{Karrai}.
Indeed, to be consistent with these results we need to include in our model a magnetic dipole $\mathbf{M}_\omega$ aligned
along the $y$ direction and such that
$\mathbf{M}_\omega=2\mathbf{e}_z\times \mathbf{P}_\omega$
where $\mathbf{e}_z$ is the direction of propagation of light
along the fiber tip~\cite{Drezet2000b,Drezet2000c}. This magnetic dipole is automatically accounted for in the model by using the distributions:
\begin{eqnarray}
\gamma_\omega(\mathbf{r})=2\sigma_0\sin{(\phi)}\delta(\rho-a)\delta(z)\nonumber \\
K_{\phi,\omega}(\mathbf{r})=-i2\omega
a\sigma_0\cos{(\phi)}\delta(\rho-a)\delta(z).
\end{eqnarray}
Finally, we point out in agreement with the result by Oberm\"{u}ller \emph{et al.}~\cite{Karrai,Drezet2000b,Drezet2000c} that the dipoles used in our model have no components in the direction perpendicular to the aperture plane. This is qualitatively different from the Bethe-Bouwkamp model~\cite{Bethe,Bouwkamp} which predicts an electric dipole along the axis normal to the aperture plane. This difference is justified on a symmetry ground due to the conical nature of the NSOM metal coating~\cite{Drezet2000b,Drezet2000c} which contrasts  with the planar geometry used in the work of Bethe and Bouwkamp.
\section{Comparing a point-like emitter to an aperture NSOM Tip}
In order to compare the spatial resolution offered by a NSOM aperture tip with the one given by a point-like dipole tip (i.e., the nanodiamond based active optical probe \cite{Cuche}) we
\begin{figure}[h]
   \begin{center}
   \begin{tabular}{c}
   \includegraphics[width=8.2cm]{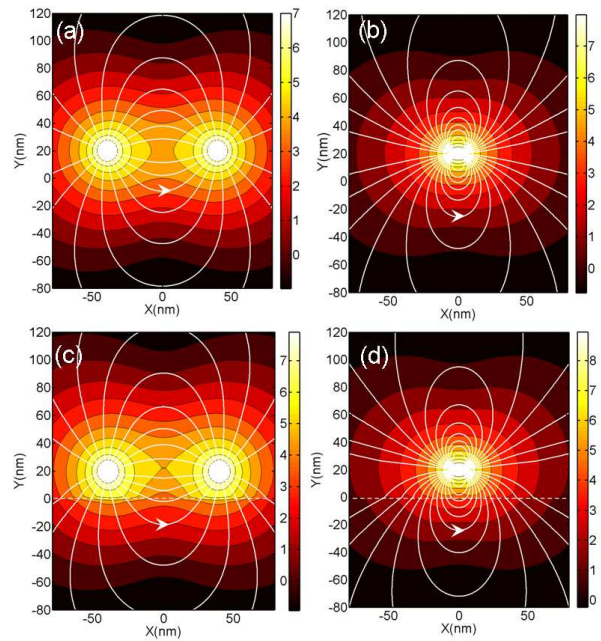}
   \end{tabular}
   \end{center}
   \caption{Electric field generated by an aperture tip, (a) and (c), and a point-like tip,  (b) and (d). (a) and (b) correspond to a free tip in vacuum while (c) and (d) correspond to the tip in front of a glass substrate. The vertical distance $h$ between both tips and the glass air interface is $h=20$ nm. The ring radius is $a=40$ nm. For each panel the electric field lines and the iso-density curves of the electric energy density $|\mathbf{E}|^2$ (in logarithmic scale) are calculated. The illumination wavelength is $\lambda=600$ nm.       }
   \end{figure}

\begin{figure}
   \begin{center}
   \begin{tabular}{c}
   \includegraphics[width=8.2cm]{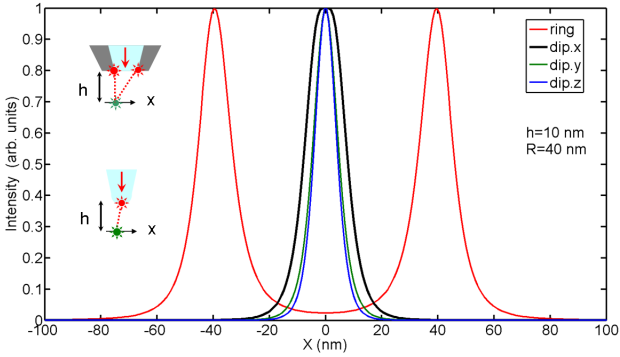}
   \end{tabular}
   \end{center}
   \caption{Simulations of the optical image obtained by scanning a fluorescent isotropical
emitter at a constant height $h=10$ nm below the NSOM tip in the ring-like and point-like configurations, respectively (see inserts). The wavelength of illumination is $\lambda=600$ nm. The black curve is the theoretical
result obtained with a dipolar point-like source (like the NV center) with a dipole oriented  along the $x$ direction. Similarly the green and blue curves are the same images for a point-like dipole along the $y$ and $z$ directions respectively. These curves are compared with
the image obtained with an usual aperture NSOM, hole radius: 40 nm  (red curve).}
\end{figure}
first calculate the field generated by both probes. Fig.~1 shows a comparison of the electric near-field
generated by the ring-like distribution (aperture radius $a=40$ nm, polarization along the $x$ axis) on
the one hand and the point-like dipolar source (dipole along the $x$ axis) on the other hand. The comparison is made for a
tip in vacuum and a tip facing a glass substrate (permittivity $\epsilon=2.25$) at a height of $h=20$ nm, respectively.
In this last configuration the reflected and transmitted fields are calculated with the image method which is known to give
consistent results in the near-field zone (see Appendix). The field generated by the ring-like distribution contains both the
electric and magnetic contributions but since $ka$ and $kh$ are much smaller than unity  we checked that the effect of the magnetic
as well as the propagative terms arising from the field propagator have negligible effects (the same is true for the propagative terms
generated by the point-like dipole). For completeness we however keep all terms in our calculations.\\
\begin{figure}[h]
   \begin{center}
   \begin{tabular}{c}
   \includegraphics[width=8.2cm]{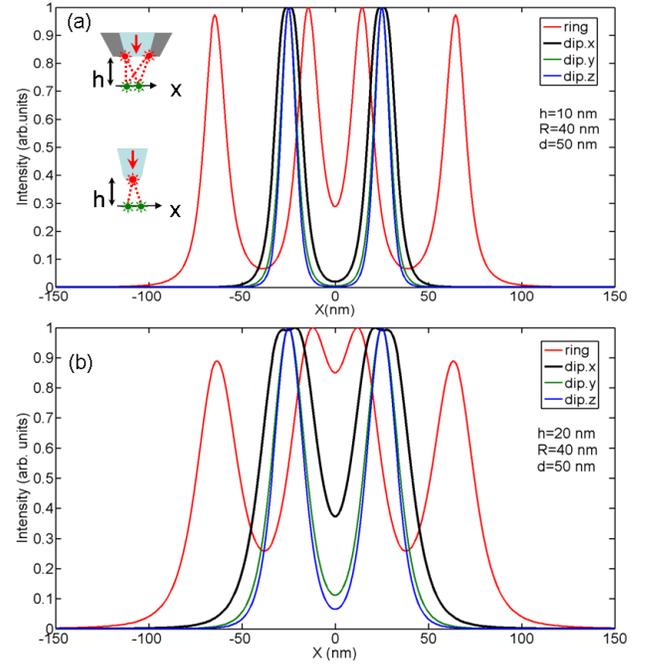}
   \end{tabular}
   \end{center}
   \caption{Simulations of the optical image obtained by scanning two fluorescent isotropical
emitters separated by a distance $d=50$ nm (in the scan direction $x$) at a constant height $h=10$ nm (panel a) or $h=20$ nm (panel b) below the NSOM tip in the ring-like and point-like configurations, respectively (see inserts). The illumination wavelength is $\lambda=600$ nm.  The black curve is the theoretical
result obtained with a dipolar point-like source (like the NV center) with a dipole oriented  along the $x$ direction. Similarly the green and blue curves are the same images for a point-like dipole along the $y$ and $z$ direction, respectively. These curves are compared with
the image obtained with an usual aperture NSOM, hole radius: 40 nm (red curve).}
   \end{figure}In a second stage, we simulate an image acquisition scan over an idealized sample. The sample is made of
either one or two point-like emitters located at the glass-air interface. To simplify, we also suppose that the emitters are
fluorescent particles emitting incoherently. The detection of the fluorescent light through the substrate is done with a
collection set-up (e.g., a microscope objective with high numerical aperture).
More precisely, we consider
the photon absorption process by a nanosphere located near
$\mathbf{r}_0$ and containing an isotropic distribution of
fluorescent emitters. Following Glauber theory~\cite{Glauber}
each emitter is excited by the field created by the tip with a
probability proportional to
$|\mathbf{E}(\mathbf{r},t)\cdot\mathbf{n}|^2$ where
$\mathbf{n}$ is the direction of the transition dipole associated
with the point-like fluorescent emitter located at $\mathbf{r}$.
Here we suppose a two-step process where the absorption is
followed by a fluorescence emission with probability
$\eta(\omega')$  at the emission frequency $\omega'$. After
averaging on the direction  $\mathbf{n}$ we therefore get a total
fluorescence signal for the nanosphere proportional to
$\eta(\omega')\cdot|\mathbf{\mathbb{E}}(\mathbf{r},t)|^2$
~\cite{Drezet2003a,Drezet2003b}. Remark that in the case of the single photon tip this picture is very
similar to the one used for describing F\"{o}rster (or
fluorescence) resonance energy transfer (FRET) between two
molecules~\cite{Novotny:2006} (we however neglect the back action
of the molecular detectors on the dynamics of the scanning
dipole).\\
Therefore the signal recorded at each tip position
is supposed to be proportional to the sum of the electric energy density $|\mathbf{E}|^2$ at the location of the
point-like fluorescent particles \cite{GerardCdF}. The collection efficiency of the NSOM microscope as used in Ref.~\cite{Cuche}
is defined by the properties of the high numerical-aperture  objective and by the numerical aperture of the multimode fiber
which guides the collected light to the detector. We estimate that 92\% of the $2\pi$ solid angle in which light is emitted
in the substrate is then collected by the optical setup. This justifies our assumption that all polarisation components of
the electric field contribute to the optical signal and therefore that this signal is proportional to $|\mathbf{E}|^2$.
We point out that in Ref.~\cite{Cuche} the $E_z^2$ contribution was neglected. We stress that using such an isotropic detector simplifies the discussion of resolution with NSOM.  Indeed, ones often consider imaging over a single molecular dipole as a genuine test of resolution. However, such a single molecule corresponds to a vectorial detector recording only information along a particular and fixed direction. While this approach was successfully used in the past for determining the orientation of molecular dipoles using an aperture  it also in general leads to more complex images in the near-field. We will therefore limit our analysis to isotropic test particles.\\
Fig.~2 shows the variation of the optical signal during a scan
along $x$ for only one isotropical emitter on top of the
substrate. The comparison between both tips reveals important
optical artifacts with the ring-like NSOM tip due to the finite
size of the ring and to the high field intensity in the rim
vicinity. These images can easily be interpreted if we consider
the fluorescent particle as a test object moving in the near-field
of the tips and scanning the emission intensity profile in a plane
at constant height $h$ above the apex. The two peaks observed
with the usual aperture NSOM are well documented in the literature
\cite{Betzig,Hulst,Drezet2003a} and are reminiscent of the high
field existing in the rim vicinity. The point-like probe does not
show such ``doubling'' of the
imaged structure and this eventually would lead to a simpler interpretation of the optical images.\\
The difference in the optical behavior between the two probes is more easily seen if we scan bothtips
over two isotropical emitters separated by a distance $d=50$ nm along the $x$ direction (see Fig.~3).
\begin{figure}[h]
\begin{center}
\begin{tabular}{c}
\includegraphics[width=8.2cm]{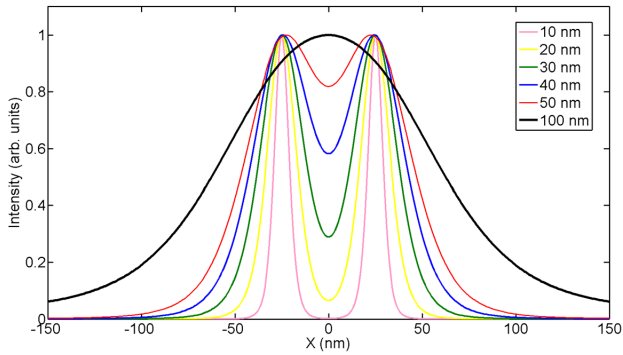}
\end{tabular}
\end{center}
\caption{Simulations of the optical image obtained by scanning two fluorescent isotropic
emitters separated by a distance $d=50$ nm (in the scan direction $x$) at a constant height $h$ below the NSOM point-like probe (i.e., the NV active tip). The different curves correspond to different $h$ going from 10 nm to 100 nm. The probe dipole is here chosen vertical ($z$ direction). }
\end{figure}Now, due to the optical artifacts, the image with the ring-like NSOM becomes much more difficult
to interpret resulting in a decrease of the
optical resolution. The effect of optical artifacts even worsens for increasing $h$,
which means that the resolving power of the aperture-NSOM probe is fundamentally limited by
both the radius $a$ and the height $h$ (compare Fig.~3a and 3b). In contrast
we see that the point-like dipole tip leads to well isolated spots during the scan. Interestingly,
the direction in which the dipole is polarized does not really change the intensity profile.
This point is of importance for practical applications since  the NV-center transition dipole is
randomly oriented (but in a fixed direction) in the nanodiamond crystal glued at the fiber tip apex~\cite{Cuche}.\\
In order to study more precisely the influence of $h$ we now
calculate the intensity profile corresponding to the scan over the
two previous isotropic emitters separated by a gap $d=50$ nm
along the $x$ direction for various heights $h$ varying in the
range 10-100 nm. The results shown in Fig.~4 for a dipole
orientation along $z$ demonstrate  clearly that the resolving
power of this kind of microscope is ultimately limited by the
height $h$ only. Here, the system offers good resolutions even for
$h=50$ nm i.e. for $h\simeq d$. Only if $h\gtrsim d $ the
resolution will be dramatically affected. These results will
actually be very general for gaps $d$ much smaller than the
wavelength since the near-field of the probe (which is essentially
wavelength independent) dominates in this spatial range.
\section{Conclusion}
In this work, we have compared theoretically the spatial resolution offered by an aperture NSOM probe with that obtained using the single-NV-center based probe. We have proposed simple analytical models for the electromagnetic field generated by both types of probe and have shown that the aperture case can be well reproduced using a ring-like current and charge distributions whereas the NV-center tip can be mimicked by a single point-like electric dipole. The electromagnetic field generated by the single NV-based tip has also been analyzed using quantum electrodynamics applied to a two-level fluorescent system and the results agree with the classical model. We have finally studied the optical signal acquisition performed during a scan over simple isotropic fluorescent objects and have demonstrated that the ultimate resolution offered by the standard aperture NSOM is limited by both the aperture diameter and the scan height. This is in contrast with a NV-center point-like source whose resolution is limited solely by the scan height. Therefore, our work stresses the importance of this new optical approach for microscopy and quantum optics in the near-field regime.
\section{Acknowledgments}
 The PhD grant of Aur\'{e}lien Cuche by the R\'{e}gion Rh\^{o}ne-Alpes (`Cluster MicroNano') is gratefully acknowledged. This work was
supported by Agence Nationale de la Recherche, France, through the NAPHO and PlasTips projects.
\section{Appendix}

The image field for an elementary (i.e. point-like) electric dipole located at position $\mathbf{r}_0=[x_0,y_0,z_0=h]$ and oriented parallel to the interface is given by:
\begin{equation}\left\{
\begin{array}{ll}
\mathbf{E}^>(\mathbf{r},t)=\mathbf{G}^>_\omega(\mathbf{r},\mathbf{r}_0)\cdot
\mathbf{P}_\omega e^{-i\omega t}\nonumber\\\simeq(\mathbf{G}^0_\omega(\mathbf{r},\mathbf{r}_0)-\frac{\epsilon-1}{\epsilon+1}\mathbf{G}^0_\omega(\mathbf{r},\mathbf{r}'_0))\cdot
\mathbf{P}_\omega e^{-i\omega t}& \textrm{for $z\geq$0} \\
&\\
\mathbf{E}^<(\mathbf{r},t)=\mathbf{G}^<_\omega(\mathbf{r},\mathbf{r}_0)\cdot
\mathbf{P}_\omega e^{-i\omega t}\nonumber\\\simeq\frac{2}{\epsilon+1}\mathbf{G}^0_\omega(\mathbf{r},\mathbf{r}_0)\cdot
\mathbf{P}_\omega e^{-i\omega t}&\textrm{for $z\leq$0}
\end{array}\right.
\end{equation} where  $\mathbf{r}'_0=[x_0,y_0,-h]$ is the position of the image dipole in the substrate medium ($z<0$, $h>0$) with dielectric permittivity $\epsilon(\omega)$.  The image method is rigorously valid only in the near-field (exception exists for the perfectly conducting metal). The case corresponding to a vertical dipole $\mathbf{P}=P_z\mathbf{e}_z$ is obtained by the substitution $\epsilon-1\rightarrow 1-\epsilon$ in the mathematical expression for $\mathbf{E}^>$ (the rest being unchanged).\\
The field generated by an elementary magnetic dipole can easily be calculated by considering a general distribution of electric  $\mathbf{J}_e$ and magnetic $\mathbf{J}_m$ current.  With such a distribution we have indeed
\begin{eqnarray}
\boldsymbol{\nabla}\times\boldsymbol{\nabla}\times\mathbf{E}-k^2\mathbf{E}=\frac{i\omega}{c}\frac{\mathbf{J}_e}{c}-\boldsymbol{\nabla}\times\frac{\mathbf{J}_m}{c}.
\end{eqnarray}
Using the definition of the dyadic Green function
\begin{eqnarray}
\boldsymbol{\nabla}\times\boldsymbol{\nabla}\times\mathbf{G}_\omega-k^2\mathbf{G}_\omega=k^2\mathbf{I}\delta^3(\mathbf{r}-\mathbf{r}_0),
\end{eqnarray} we deduce the integral formula
\begin{eqnarray}
\mathbf{E}(\mathbf{r},t)=\int
\mathbf{G}_\omega(\mathbf{r},\mathbf{r}')\cdot
[\frac{i\mathbf{J}_e(\mathbf{r}')}{\omega}-c\boldsymbol{\nabla}\times\frac{\mathbf{J}_m}{\omega^2}]d^3\mathbf{r}'e^{-i\omega t}.
\end{eqnarray}In particular for a point-like magnetic dipole  $\mathbf{J}_m=-i\omega \mathbf{M}\delta^3(\mathbf{r}-\mathbf{r}_0)$ located above the interface one gets after integration by parts
\begin{eqnarray}
\mathbf{E}^{>,<}(\mathbf{r},t)=\frac{i}{k}[\boldsymbol{\nabla}\times\mathbf{G}^{>,<}_\omega(\mathbf{r},\mathbf{r}_0)]\cdot\mathbf{M} e^{-i\omega t}.
\end{eqnarray}


\begin{thebibliography}{99}
\bibitem{foot}
The Heisenberg formula~\cite{Heisenberg} Eq.~10 is actually an
oversimplification of the problem as shown by Bohr and
Rosenfeld~\cite{Bohr}.  Indeed, a rigorous quantum calculation
shows that $[\overline{E}_i,\overline{B}_j]=0$ (i.e., $\delta
\overline{E}_i\cdot\delta \overline{B}_j \geq 0$). Consequently
there is no limit to simultaneous measurements of $\overline{E}_i$
and $\overline{B}_j$. It is true however, that the field operators
defined at two space-time locations $a$ and $b$ (i.e. not averaged
on a finite volume) do not in general commute:
$[E_i(a),B_j(b)]\neq 0$~\cite{Bohr}. Therefore, the problem
discussed in \cite{Barnes, Dereux} concerning the impact of
quantum optics on NSOM resolution is still relevant.
\bibitem{remarque}
This adiabatic approximation is valid for moderate saturation: $\Omega_R^2\ll (\Gamma/2)^2+(\omega_L-\omega_{eg})^2$ \cite{Tannoudji}.
\end{thebibliography}
\end{document}